\documentclass[aps,twocolumn,superscriptaddress,amsmath]{revtex4}

\usepackage[dvips]{graphicx}
\usepackage{dcolumn}
\usepackage{amsmath}
\usepackage{threeparttable}

\usepackage{bm}
\usepackage[version=3]{mhchem} 

\newcommand{\be}{\begin{equation}}
\newcommand{\ee}{\end{equation}}
\newcommand{\bea}{\begin{eqnarray}}
\newcommand{\eea}{\end{eqnarray}}
\newcommand{\bean}{\begin{eqnarray*}}
\newcommand{\eean}{\end{eqnarray*}}

\begin{document}


\title{Resonating valence bond wave function with 
molecular orbitals: Application to first-row molecules}

\author{Mariapia Marchi}
\email{marchi@sissa.it}
\affiliation{ SISSA, International School for Advanced Studies, 34151, Trieste, Italy}
\affiliation{DEMOCRITOS National Simulation Center, 34151, Trieste, Italy}

\author{Sam Azadi}
\email{azadi@sissa.it}
\affiliation{ SISSA, International School for Advanced Studies, 34151, Trieste, Italy}
\author{Michele Casula}
\email{michele.casula@gmail.com}
\affiliation{Centre de Physique Th\'eorique, Ecole Polytechnique, CNRS, 91128 Palaiseau, France}
\author{Sandro Sorella}
\email{sorella@sissa.it}
\affiliation{ SISSA, International School for Advanced Studies, 34151, Trieste, Italy}
\affiliation{DEMOCRITOS National Simulation Center, 34151, Trieste, Italy}

\date{\today}

\begin{abstract}
We introduce a method for accurate quantum chemical 
calculations based on a simple variational wave function,  
defined by a single geminal   
that couples  all the electrons into singlet pairs,
combined with a real space correlation factor.
The method uses a constrained variational optimization,
based on an  expansion of   the geminal  in 
terms of molecular orbitals.
It is shown that the most relevant non-dynamical correlations 
are correctly reproduced once an appropriate number $n$ 
of molecular orbitals is considered.
The value of $n$ is  determined by 
requiring that, in the atomization limit, the atoms are described
by Hartree-Fock Slater determinants with Jastrow 
correlations.    
The energetics, as well as other physical and chemical 
properties, are then given
by an efficient variational approach based on  
standard  quantum Monte Carlo techniques. 
We test this method on a set of homonuclear 
(\ce{Be2}, \ce{B2}, \ce{C2}, \ce{N2}, \ce{O2}, and \ce{F2}) and heteronuclear
(\ce{LiF}, and \ce{CN}) dimers 
for which strong non-dynamical correlations and/or 
weak van der Waals interactions are present. 
\end{abstract}

\maketitle

\section{Introduction}
\label{introduction}

Already in the early stages of quantum mechanics, 
L. Pauling introduced the so called resonating valence bond (RVB)
theory of the chemical bond\cite{pauling}, starting from the simple
consideration that a spin singlet can be formed between any two valence 
electrons belonging to neighboring atoms.
In this scheme, the ground state wave function of a molecule,
such as benzene, can lower the energy by allowing the resonance among all 
possible valence bond configurations that can be drawn by linking the 
positions of two atoms (e.g. the Kekul\'e and Dewar configurations in the 
benzene molecule).
However, its application was limited, since 
the number of bonds were growing exponentially with the number of atoms. 
As a consequence, the powerful language of molecular orbitals (MO's) applied to 
Hartree-Fock (HF) and post HF methods became popular. 
Nonetheless, quite recently, the interest in RVB wave functions has been 
strongly revived. 
Indeed, soon after the discovery of the High-Tc superconductors, 
P. W. Anderson realized that a single determinant wave function 
combined with a suitable real space correlation term -- henceforth 
referred to as 'the Jastrow factor' --
could be used to represent a complex RVB state.\cite{pwanderson}
In this new ansatz a crucial ingredient is the form of 
the determinantal part of the wave function,  that is required to be 
a singlet state with total spin $S=0$.
This picture, aimed at
explaining the High-Tc superconductivity, represents also a very 
efficient numerical implementation of the original RVB idea, soon 
reconsidered in this form for lattice models\cite{gros,dagotto_rev,dagotto_tj_sup,ogata,spanu_tj}, 
and then in realistic simulations of atoms and small molecules\cite{casula1,casulamol,pfaff1,pfaff2}.
Though the Anderson's RVB wave function has been originally defined just for singlet states, 
the same concept can be applied to electronic systems with arbitrary spin $S>0$,
with the inclusion of unpaired orbitals. 
This is a very important generalization in order to describe polarized compounds,
like the transition element compounds which show high-spin configurations in their low-lying energy states.
In the actual RVB description of realistic systems,
it is necessary to resort to standard 
quantum Monte Carlo (QMC) methods\cite{mitas} in order to compute the 
variational expectation values of the energy 
and correlation functions\cite{casula1,casulamol,rocca}.

In this article, we propose an extension of the RVB picture that  
is  based on a MO expansion of the singlet valence bond pairs defining the wave 
function. This ansatz yields a correlation consistent 
RVB representation by means of a constrained energy minimization which keeps 
the number of MO's fixed while stretching the bond. 
By setting this number to a value such that 
a Jastrow correlated HF wave function is recovered in the atomization limit, 
we obtain, with a single determinant, a remarkably accurate description of the 
bond, even when strong non-dynamical correlations are present in the system.
In this paper we illustrate the method and test it on a set of dimers composed
by first row atoms and on selected small molecules belonging to the so-called 
'G1 set' (see Ref.~\cite{grossman}), often used to test new 
theoretical methods. 
The approach described in this work has also been applied to 
the study of the controversial ground state of the iron dimer\cite{iron2}.

In the following, we describe the RVB wave function and our extension, and we 
show test results on various homonuclear and heteronuclear dimers
(\ce{Be2}, \ce{B2}, \ce{C2}, \ce{N2}, \ce{O2}, \ce{F2}, \ce{LiF}, and \ce{CN}). 
In Appendix \ref{molorb} we describe the constrained minimization of
the molecular orbital expansion of the trial wave function. 
In Appendix~\ref{check_app}, we present a  systematic study 
of the  variational energies obtained with the RVB wave function, 
as a function of the  number of molecular orbitals 
(Appendix~\ref{c2molapp}) and the size of the atomic  basis set used
(Appendix~\ref{basisconvergence}).

\section{Variational method}

\subsection{General description of the wave function}
The fundamental ingredient of our variational method 
is an N-electron RVB wave function, called JAGP since it is
the product of a Jastrow factor J, and a determinantal part which is
an antisymmetrized geminal power (AGP), previously 
introduced in Refs. \cite{casula1,casulamol}
($\Psi_{JAGP}=J\,\Phi_N$).
Below we shall describe this wave function.

In the case of $N$ electrons with $N_\uparrow$ up spins  
($N_\downarrow$ down spins),  
where for simplicity we take $N_\uparrow\ge N_\downarrow$, we
can describe a pure spin state with total spin 
$S=|N_\uparrow-N_\downarrow|/2$ and maximum spin projection 
$S^{tot}_z=S$ by means of the antisymmetrized product
of $N_\downarrow$ singlet pairs and $2S$ unpaired orbitals 
corresponding to the remaining spin-up electrons. 
Hence the determinantal part reads 
\begin{equation} 
\label{AGP}
\Phi_N(\vec R)= {\cal A }  \prod\limits_{i=1}^{N_\downarrow} 
\phi (\vec r^\uparrow_i, \vec r^\downarrow_i ) 
\prod_{j=N_\downarrow+1}^{N_\uparrow} \phi_j (\vec r^\uparrow_j ),
\end{equation}
with ${\cal A}$ the antisymmetrization operator, 
$\vec R= \left\{ \vec r_1^\uparrow, 
\cdots, \vec r_{N_\uparrow}^\uparrow, r_1^\downarrow, \cdots ,  
\vec r_{N_\downarrow}^\downarrow \right\}$ the $3N$-dimensional
vector of coordinates, 
$\phi(\vec r,\vec r^\prime) =\phi(\vec r^\prime,\vec r)$ 
a symmetric orbital function describing the singlet pairs, and 
$\phi_j(\vec r)$ the unpaired orbitals.
It can be shown that the wave function in Eq.~(\ref{AGP}) can 
be rewritten in terms of a single determinant 
(see Ref.~\cite{casula1} and references therein).

$\Phi_N(\vec R)$ has definite total spin. We also impose all 
possible symmetries to be satisfied by Eq. (\ref{AGP}), such 
as angular momentum and spatial reflections.

Similar constructions with definite spin 
can be done, by allowing also triplet pairing 
between the $2S$ unpaired electrons.
Since this involves a bit more complicated algebra like the use of 
pfaffians\cite{pfaff1,pfaff2}, we do not consider it here.

The Jastrow factor 
takes into account the electronic correlation between two 
electrons and 
 is conventionally 
split into a homogeneous interaction $J_2$ depending on the relative 
distance between two electrons
(i.e. a two-body term),  and a non homogeneous 
contribution depending also on the electron-ion distance, 
included in the one-body $J_1$, three-body $J_3$ and four-body $J_4$ terms.
$J_1$ is a single particle function which is important to compensate the 
change in the one particle density induced by $J_2$, $J_3$ and $J_4$, 
as well as to satisfy the electron-ion cusp conditions.
The one- and two-body terms $J_1$ and $J_2$ are defined by the following 
equations: 
\begin{eqnarray}
\label{j1}
J_1=\exp{\big[\sum_{ia}-(2Z_a)^{3/4}u(Z_a^{1/4}r_{ia})+\sum_{ial}
g_l^a \chi_{al}^J(\vec{r}_i)\big]},
\end{eqnarray}
and 
\begin{eqnarray}
\label{j2}
J_2=\exp{[\sum_{i<j}^{}u(r_{ij})]},
\end{eqnarray}
where $i,j$ are indexes running over the electrons, and $l$ 
runs over different single particle orbitals $\chi_{al}^J$ 
centered on the atomic center $a$.
$r_{ia}$ and $r_{ij}$ denote electron-ion and 
electron-electron distances respectively. 
The corresponding cusp conditions are fixed by 
the function $u(r)=F[1-\exp(-r/F)]/2$ (see e.g. 
Ref.~\cite{rocca}).
$g_l^a$ and $F$ are optimizable variational parameters.

The three- and four-body Jastrow $J_{3} J_{4}$ are  given by:
\begin{equation}
\label{jastrow}
J_{3} J_{4}(\vec R)=
\exp\left(\sum \limits_{i<j} f(\vec r_i ,\vec r_j)\right), 
\end{equation}
with $f (\vec r ,\vec r^\prime)$, being  a two-electron coordinate 
function that  can be expanded into the same single-particle basis 
used for $J_{1}$: 
\begin{eqnarray}\label{3bjsp}
f(\vec{r}_i,\vec{r}_j)=
\sum_{ablm}^{} g_{lm}^{ab}\,\chi_{al}^{J}(\vec{r}_i)
\chi_{bm}^{J}(\vec{r}_j),
\end{eqnarray}
with $g_{lm}^{ab}$ optimizable parameters. 
Three-body (electron ion electron) 
correlations are described by the diagonal matrix elements 
$g^{aa}$, 
whereas four-body correlations (electron ion electron ion) are
described by matrix elements with $a\ne b$.

The  exhaustive and complete expression of the  
Jastrow factor  $J(\vec R) = J_1(R) J_2 (R) J_3(R) J_4(R) $ 
that we adopt in this work allows to take 
into account not only weak   
electron-electron interactions of the van der Waals (vdW) type, 
but it is also extremely effective for 
suppressing  higher energy configurations
with overlapping valence bonds, which otherwise 
lead to a too large electron density around an atom. 

As any functions of two coordinates, also the pairing function
$\phi$ in Eq.~(\ref{AGP}) can be 
expanded in terms of single particle orbitals. We can thus 
write
\begin{equation} \label{pairing}
\phi( \vec r, \vec r^\prime) = \sum\limits_{j=1}^{n-2S} 
\lambda_j \phi_j( \vec r) \phi_j( \vec r^\prime), 
\end{equation}
where $n$ is large enough, and $\{ \phi_j \}$ is an orthogonal 
single particle basis set, which reaches 
its complete basis set limit (CBS) for $n\to \infty$.
Notice that, in these notations, we assume that the $2S$ 
unpaired orbitals $\phi_j$ of  Eq.~(\ref{AGP}) 
correspond to the indexes $ n-2S+1 \le j \le n$ 
in Eq.~(\ref{pairing}).

The single particle orbitals $\phi_j$ can be conveniently 
chosen as the MO's obtained with a conventional restricted HF 
(RHF) calculation. The MO basis allows us to write 
Eq.~(\ref{pairing}) in a diagonal form equivalent to a more 
involved matrix form when the 
MO's are  developed in an atomic basis set \cite{casulamol} of 
orbitals $\varphi_{a,j}$ where $a$ indicates the atomic center 
and $j$ the type: 
$\phi_i (\vec r) =\sum_{a,j}\zeta^i_{a,j}\varphi_{a,j}(\vec r)$.
The coefficients $\zeta^i_{a,j}$, as well as the weights 
$\lambda_j$, 
can be used as variational parameters defining the geminal in 
Eq.~(\ref{pairing}).
By truncating the expansion in Eq.~(\ref{pairing}) 
to a number of MO's $n$ equal to the number of electron pairs 
plus the unpaired orbitals, namely $n=N_\uparrow$, one recovers 
the usual RHF theory, because the antisymmetrization operator 
${\cal A}$ singles out only 
one Slater determinant. Moreover, the MO weights  $\lambda_j$ 
affect only an overall prefactor of this Slater determinant, 
so that their actual values are irrelevant in this case. 
However, the pairing function is generally 
not limited to have only $N_\downarrow$ non vanishing 
eigenvalues $\lambda_j$.
Therefore, the RVB wave function represents a clear extension 
of the RHF theory, not only for the  presence of the Jastrow 
factor, which 
considerably improves the dynamical correlations, but mainly  
because its determinantal part goes beyond RHF 
when $n>N_\uparrow$, by including also non-dynamical 
correlations.
Quite generally, a gain in energy and a more accurate 
calculation are expected whenever $n> N_{\uparrow}$. 

\subsection{Valence bond energy 
consistent number of molecular orbitals in the AGP}
\label{molnumber}
The main property used in the following derivation
relies on the fact that the atoms are well described 
by a Jastrow correlated RHF (JHF) wave function. Indeed, the 
application of the Jastrow factor $J$ on a simple 
HF Slater determinant
provides at least $\simeq 90\%$ of the correlation energy in all 
the atoms (see Refs.~\cite{mitas,casula1} and Table~\ref{tab_jhf_at}).
Here we show that it is possible to extend and remarkably improve
the correlated HF approximation for the chemical bond,
by means of a RVB wave function with an \emph{appropriate} number $n$
of MO's appearing in Eq.~(\ref{pairing}). These MO's are
chosen to minimize the energy expectation
value in presence of the Jastrow factor,
while an upper bound on the number $n$ is univocally determined by imposing 
that,
when the atoms are at large distance, we cannot obtain  an energy
below the sum of the JHF atomic energies.
\begin{table*}
\begin{center}
\begin{threeparttable}
\caption{\small{
Atomic energies for \ce{Li}, \ce{Be}, \ce{B}, \ce{C}, \ce{N}, \ce{O}, and
\ce{F}: comparison between RHF benchmarks, estimated exact values, VMC and LRDMC
JHF data, and the percentage of recovered correlation energy (\%) (evaluated
using the estimated exact value and the Hartree-Fock energy).  
For \ce{Li} and \ce{Be} all-electron results are shown. For all the other atoms, 
results were obtained with a pseudopotential (Ref.~\cite{filippipseudo}). 
}}
\label{tab_jhf_at}
\begin{tabular}{| c| c| c| c| c| c| }
\hline\hline
 Atom & RHF & Est. exact & JHF VMC & JHF LRDMC & \% \\
\hline
\ce{Li} & -7.432727\tnote{a}  & -7.47806\tnote{a} & -7.47707(6) &-7.47807(3)  & 100\% \\
\ce{Be} & -14.573023\tnote{a} & -14.66736\tnote{a}  & -14.64747(9) &-14.6575(1)  & 89.5\% \\
\ce{B} & -2.54375616\tnote{b} & -2.61940948\tnote{b} & -2.6031(1) & -2.6110(1) & 88.9\%\\
\ce{C} & -5.32903005\tnote{b}& -5.43249352\tnote{b} & -5.4105(1) & -5.4216(1) & 89.5\%\\
\ce{N} &-9.66837630\tnote{b} & -9.79973109\tnote{b} & -9.7771(3) & -9.7898(1) & 92.4\% \\
\ce{O} &-15.70844748\tnote{b} & -15.90165954\tnote{b} & -15.8754(1) & -15.89233(8) & 95.2\% \\
\ce{F} & -23.93849161\tnote{b}& -24.19290003\tnote{b} & -24.1680(3)  & -24.1860(2)  & 97.3\%\\
\hline
\hline\hline
\end{tabular}

\begin{tablenotes}
{\footnotesize
\item[a]{From Ref.~\cite{hfatoms}}
\item[b]{From Ref.~\cite{filippipseudo}}
}
\end{tablenotes}
\end{threeparttable}
\end{center}
\end{table*}

The above mentioned criteria are based on the assumption
that the large intra-atomic correlations do not affect the 
chemical properties of the bond, which are instead extremely 
sensitive to the usually much weaker inter-atomic 
correlations. Moreover,
electrons close to the atomic centers
are chemically inert because they are far away from the region
where the bond is formed. 
Hence, an improvement in the description of the atoms with 
many determinants\cite{umrigar} would lead in this case only 
to a rigid shift of the total energy. 
The above assumption is a quite generally accepted idea, 
that has been exploited in different ways by a large variety of approaches. 
For instance, it validates  the use of pseudopotentials, 
the configuration interaction (CI) with the frozen core 
approximation\cite{fci},
and is the basis for other quantum chemistry methods such 
as the symmetry-adapted perturbation theory (SAPT)\cite{sapt},
and the Morokuma analysis 
(see Ref.~\cite{hfsizeconsistent} and references therein).

In the following we shall denote the aforementioned
\emph{appropriate} number $n$ of MO's with $n^*$.
Let us denote with $M$ a molecule composed by atoms $A_1$, 
$A_2$, etc. 
The optimal value of $n^*$ is most generally obtained by 
saturating a simple  upper bound value $\tilde n$: 
\begin{eqnarray}\label{ntilde}
n^* \le \tilde n=\sum_{i} N_{\uparrow} (A_i),
\end {eqnarray}
where $i$ is an index running over the atoms composing the 
molecule  $M$.
Since in some cases convergence in the energy for the JAGP can be 
obtained even for  $n^*< \tilde n$, we have used the inequality 
to define $n^*$ and the corresponding wave function will be denoted 
by JAGPn$^*$. 
If the sum of the number of spin-up 
electrons in the atoms equals the number of the MO's required
by a RHF calculation $n_{HF}(M)$ for the molecule, then
$n^*=n_{HF}(M)$ and the JAGPn$^*$ wave function reduces to a
JHF description of the molecule. 
This is the case, e.g., for \ce{Be2} and \ce{B2}.
In all the other cases we have 
$n_{HF}(M)< \tilde n$, and, in this work, 
we have found that
there is a substantial energy gain in increasing the number of 
molecular orbitals with respect to the RHF value. This happens for 
instance for \ce{N_2}, \ce{O_2},  
\ce{F_2}, and \ce{CN},  whereas for \ce{LiF}, 
though $\tilde n> n_{HF}$, accurate results can be obtained even with 
$n^*=n_{HF}$.

The upper bound in  Eq.~(\ref{ntilde}) can be slightly improved, 
as it will be shown in the following.
This is particularly important 
when some degenerate multiplets of 
orbitals are not completely occupied, as for the \ce{C2} molecule where, 
by using $\tilde n$ molecular orbitals in the AGP expansion, 
one of the two antibonding $\pi^*$ orbitals remains empty,
and therefore it is not possible to satisfy the orbital 
symmetry of 
the $^1\Sigma_g^+$ \ce{C2} wave function. 
In the general case  the highest molecular orbital included in 
the AGP has degeneration $D$ and it may occur that only $\tilde D < D$ 
orbitals of the multiplet are included in the AGP expansion by the upper bound in Eq.~(\ref{ntilde}).
For this reason it is important to  improve the upper bound 
(\ref{ntilde}) for $n^*$, in particular cases 
when the chemical compound is spatially symmetric,  
namely for  reflections, rotations, translations, of the atomic positions.
In fact, let us suppose that the molecule is composed by several  atoms. 
Some spatial symmetry operations can make equivalent $n_A\ge 1$ 
identical atoms of type $A$.  
Assuming that these  symmetries remain valid  up to the atomization limit, 
we denote by $m$ the minimum value of $n_A$ among all atomic 
species. Then if $m>1$ 
it  is possible to improve the upper bound (\ref{ntilde})  by:  
 \begin{equation} \label{therightn}
 n^* \le \tilde n +m-1.
 \end{equation}
For instance for \ce{C2}, 
according to Eq.~(\ref{therightn})  we have $m=2$ due to the reflection 
symmetry of the molecule, and  
$n^* \le \tilde n +1$. Indeed $n^* = \tilde n +1$ not only allows to fulfill
the $^1\Sigma_g^+$ symmetry, but also provides a substantial improvement of
the binding with respect to $n^* = \tilde n$ (see 
Fig.~\ref{c2mol} in Appendix~\ref{c2molapp}). 
The one extra molecular orbital added cannot have any effect at large distance 
in a fully symmetric calculation that connects the compound at
rest to $m=2$ equivalent  Carbon atoms at large distance. 
Indeed, in this case,  the presence of the extra orbital could  
improve only  the energy of one of the two JHF atoms, thus violating their equivalence.
Therefore the eigenvalue  $\lambda_{j}$ of Eq.~(\ref{pairing}) corresponding to 
the extra molecular orbital  must vanish  in the atomization limit.
 
Generally speaking  a  value for  $n$  larger than 
the upper bound (\ref{therightn}) certainly leads to a lower 
value of the 
total energy, but may improve much more the atomic energies, 
rather than the bonding.
Actually, we have seen that, in all cases so far considered, 
the accuracy in describing 
the chemical bond  improves systematically 
by increasing the number of molecular orbitals, provided it remains smaller 
than  the upper bound. 
Clearly, whenever Eqs.~(\ref{ntilde}) or (\ref{therightn}) are satisfied 
the atomization energy has to be 
referred to the JHF calculation, 
even if lower energies could be achieved with a 
JAGP wave function for the atoms\cite{casula1}.
Remarkably, in the limit of large number of molecular orbitals,  
when the lowest JAGP total energies are  obtained both for the atom and the 
molecule,
the binding energy becomes always worse than the corresponding JAGP$n^*$.

The JAGPn$^*$ wave function can be used also to describe bulk systems by 
applying  
the upper bound of Eq.~(\ref{ntilde}) and of Eq.~(\ref{therightn}) to the 
supercell containing a finite number of atoms, so that the values of 
$\tilde n$ and $m$ easily 
follow exactly as in the case of a finite open system.
The upper bound computed in this way may 
exceed by a large amount the number $n_{HF}$ of molecular 
orbitals necessary to define  a single  Slater determinant in the supercell.
Thus, convergence in the energy is expected in this case  
for $n^* \ll \tilde n$.
For instance, in the case of graphene 
for a typical supercell of 48 atoms, $\tilde n=4\times 48=192$, whereas 
$n_{HF}=3 \times 48=144 \ll 192$.

The constrained optimization of the JAGP 
wave function with a given number of MO's is 
a generalization of the standard QMC optimization\cite{umrigar}
which minimizes the total energy, and will be described in 
Appendix~\ref{molorb}.  

\section{Results}
In this section we shall describe the results that we have 
obtained for a set of molecules composed of first row atoms, 
where strong non-dynamical correlation and/or weak vdW 
interactions are present. These molecules are used as a test-case 
for our method. 

Our study has been carried out by means of QMC simulations. We started from the 
constrained optimization of the variational wave function described in the 
previous section, which was initialized by taking density functional theory  
orbitals in the local 
density approximation, and then we performed variational Monte Carlo (VMC) or 
lattice regularized diffusion Monte Carlo 
(LRDMC) simulations\cite{lrdmc}.

For the determinantal part of the wave function we have used a Slater 
(for \ce{Be2}, and the \ce{Li} atom in the \ce{LiF} molecule), or mixed 
Slater/Gaussian 
(for \ce{B2}, \ce{C2}, \ce{N2}, \ce{O2}, \ce{F2}, \ce{CN}, and the \ce{F} atom 
in the \ce{LiF} molecule) 
basis, large enough for an accuracy of 1mH in the total energies.
This quantity sets the tolerance for our complete basis set (CBS) limit 
extrapolation.
In particular, for \ce{Be2} the basis set is $6s4p2d$, for \ce{B2} $5s4p1d$,
for \ce{C2} $5s5p$, for \ce{N2} 
$5s3p2d$, for \ce{O2} $6s5p2d$, for \ce{F2} $5s5p2d$, for the \ce{Li} atom in 
the \ce{LiF} molecule $5s4p$, whereas for the \ce{F} atom, as well as for the 
\ce{C} and \ce{N} atoms composing the \ce{CN} molecule, we used the same basis 
adopted for the corresponding dimers. In the mixed 
Slater/Gaussian cases we have used one Slater orbital for each angular momentum,
except for $d$ orbitals, 
which have been chosen of a purely
Gaussian form.
Thus, by fully optimizing all the coefficients and the exponents of the 
primitive basis set, we 
have verified that the dimension of the basis is sufficient to 
achieve the desired accuracy. In Appendix~\ref{basisconvergence} we show, as an
example, selected studies of convergence in the basis set.

A much smaller basis was used for the Jastrow factor, 
because this allows for a more efficient energy optimization.
On the other hand, 
the essentially exact contribution of Jastrow-type dynamical correlations,
which do not change the phases of the wave function, can be very accurately 
obtained with the well established DMC technique\cite{mitas}, 
within the recent lattice regularized diffusion Monte Carlo (LRDMC) 
implementation\cite{lrdmc}.
LRDMC  is equivalent to standard 
DMC for all-electron calculations, and represents an 
improvement of the older technique because  it allows to obtain 
a rigorous upper bound of the total energy even when  pseudopotentials are used 
in the calculation. 
The DMC/LRDMC 
 approach can be seen as a stochastic optimization of the Jastrow factor
which keeps fixed the phases of the RVB wave function.
In some test cases (see Appendix~\ref{basisconvergence}), we have also verified 
that a larger basis in the Jastrow 
does not provide significant changes in the physical and chemical 
quantities here considered, because total energy differences are much less 
sensitive to the extension of the Jastrow basis set.

We have used a helium-core pseudopotential\cite{filippipseudo} 
for all but \ce{Be} and \ce{Li} atoms. 
In some  test cases without pseudopotentials (e.g. \ce{Be2})
we have explicitly verified  
that the DMC and the LRDMC 
energies are consistent, but we have adopted the latter method for the sake 
of generality.  
In the \ce{C2} case we have also checked that the effects
of the pseudopotential on the total energy differences are negligible\cite{c2pseudo}.

\begin{table*}
\begin{center}
\begin{threeparttable}

\caption{\small{
Bond lengths (in a.u.), well depths (in eV) and ZPE (in mH) for a set of 
first row diatomic molecules. We 
report VMC and LRDMC values for both JHF and 
JAGPn$^*$ trial wave functions, and experimental results or estimated exact 
values.
The well--depth exact estimates are given by the 
experimental binding energies subtracted
by the  spin-orbit energies when accessible 
 (i.e. for all atoms but \ce{B2} and \ce{Be2}) and the ZPE. 
We also report the J$\times$SD DMC well depths of Ref.~\cite{umrigar} 
when available.
For \ce{Be2} (all electron calculations) and \ce{B2} (calculations with the pseudopotential in Ref.~\cite{filippipseudo}), 
$n^*=N_\uparrow=4$, hence the JHF and JAGPn$^*$ results coincide. 
}}\label{tab_molecules}
\begin{tabular}{|l| c| c| c| c| c| c| c| c|}
\hline\hline
\multicolumn{9}{|c|}{{\it Bond length} (a.u.)}\\
\hline
& \ce{Be2} (all el.) & \ce{B2} & \ce{C2} &\ce{N2} &\ce{O2} &\ce{F2}&\ce{LiF}&\ce{CN} \\
\hline
JHF VMC         & 4.85(5) & 3.041(6)    &2.367(2) & 2.048(1)  & 2.27(1) & 2.66(1) &2.95(4)&2.185(6)\\
JAGPn$^*$ VMC   &    4.85(5) & 3.041(6) &2.334(6) & 2.075(2) & 2.268(7) & 2.661(5)&2.92(2)&2.200(6)\\
JHF LRDMC       & 4.65(7) & 3.021(9)    &2.369(3) & 2.051(1)  & 2.270(4) & 2.665(9)&2.949(8)&2.201(3)\\
JAGPn$^*$ LRDMC &    4.65(7)   & 3.021(9)      &2.337(6) & 2.075(1)  & 2.277(4) & 2.663(3)&2.950(7)&2.202(2)\\
Exact estim. & 4.63\tnote{a}& 3.005\tnote{b} & 2.3481\tnote{c} & 2.075\tnote{b} & 2.283\tnote{b} & 2.668\tnote{b}&2.955\tnote{b}&2.214\tnote{b}\\
\hline
\multicolumn{9}{|c|}{{\it Well depth} (eV)}\\
\hline
& \ce{Be2} (all. el) & \ce{B2} & \ce{C2} &\ce{N2} &\ce{O2} &\ce{F2}&\ce{LiF}&\ce{CN}  \\
\hline
JHF VMC         & 0.120(5) & 2.754(3)& 5.538(9)  & 9.662(3) & 4.976(8) & 1.124(4)& 5.93(2)  &7.52(1)\\
JAGPn$^*$ VMC   & 0.120(5) & 2.754(3)& 6.327(9)  & 9.874(2) & 5.060(7) & 1.671(2)& 5.96(2)  &7.68(1)\\
J$\times$SD DMC & 0.125(1) &2.798(3) & 5.656(3)  & 9.583(3) & 4.992(7) & 1.349(6)& -- &--\\
JHF LRDMC       & 0.143(6) &2.797(2) & 5.763(9)  & 9.665(2) & 5.070(5) & 1.452(3)& 6.049(6) &7.661(5)\\
JAGPn$^*$ LRDMC & 0.143(6) & 2.797(2) & 6.297(8)  & 9.882(1) & 5.126(5)  & 1.688(2)& 6.056(6) &7.744(5)\\
Exact estim. & 0.1153(3)\tnote{a} &2.91(6)\tnote{d}& 6.43(2)\tnote{e} & 9.902(3)\tnote{e}  & 5.233(3)\tnote{e}  & 1.693(5)\tnote{e}&6.03(9)\tnote{f}&7.86(9)\tnote{f} \\
\hline
\multicolumn{9}{|c|}{{\it ZPE} (mH)}\\
\hline
& \ce{Be2} (all el.) & \ce{B2} & \ce{C2} &\ce{N2} &\ce{O2} &\ce{F2}&\ce{LiF}&\ce{CN} \\
\hline
JHF VMC         & 0.56(5) &  2.49(5) & 4.3(1) & 6.38(6)  & 3.8(1) & 2.20(3)& 2.3(2) & 4.9(1) \\
JAGPn$^*$ VMC   & 0.56(5) &  2.49(5) & 4.2(1) & 5.48(3)  & 3.85(9) & 2.20(3)&2.1(2)&4.87(8)\\
JHF LRDMC       & 0.61(9) &  2.51(7) & 4.38(3)& 5.83(6)  & 3.77(5) & 2.16(3)& 2.18(8)& 4.81(3)\\
JAGPn$^*$ LRDMC & 0.61(9)  & 2.51(7) & 4.3(1) & 5.51(2)  &3.70(9)  & 2.22(2)&2.10(6)&4.82(4)\\
Exp. & 0.56\tnote{a}& 2.4\tnote{b} & 4.2\tnote{e} & 5.4\tnote{e} & 3.6\tnote{e} & 2.1\tnote{e}&2.07\tnote{b}&4.71\tnote{b}\\
\hline\hline
\end{tabular}

\begin{tablenotes}
{\scriptsize
\item[a]{From Ref.~\cite{be2science}}
\item[b]{From Ref.~\cite{huber}}
\item[c]{From Ref.~\cite{filippic2}}
\item[d]{From Ref.~\cite{bauschlicher}}
\item[e]{From Ref.~\cite{bytautas}}
\item[f]{From Ref.~\cite{feller}}
}
\end{tablenotes}

\end{threeparttable}
\end{center}
\end{table*}
In Table~\ref{tab_molecules}, we compare with estimated exact results bond lengths and 
well depths 
obtained by means of VMC and LRDMC simulations 
performed with JHF or JAGPn$^*$ wave functions for the various molecules 
considered in this paper. We optimized each wave function for a bunch of different
interatomic distances. The energy and interatomic distance at the minimum were found
by interpolating the energy close to its  minimum value with 
 a  cubic polynomial.
We also report binding energies found in Ref.~\cite{umrigar} with DMC calculations
for a fully optimized all-electron Jastrow-correlated single determinant wave
function (in our table denoted with  J$\times$SD DMC).
Finally, we compare the JAGPn$^*$ zero point energy (ZPE) with available 
experimental data. 
This quantity was computed by standard first order perturbation theory in 
the anharmonic cubic term.
 For this property, the agreement between both VMC and LRDMC
results and experimental findings is satisfactory in most of the cases.
The accuracy in the  ZPE can be probably improved by doing  a more careful fit 
around the minimum.  

Below we comment all the different cases.

\subsection{Beryllium and boron dimers}
Though the Beryllium dimer does not belong to the so-called 'G1-set' of 
molecules (see e.g. Ref.~\cite{grossman}), this dimer is a very
important test case for the variational method proposed in this paper.
Indeed, several computational methods (see e.g. Ref. \cite{milestone}), 
including previous QMC simulations\cite{casulamol}, 
have failed in the 
attempt of reproducing the binding of this molecule. Moreover, until the 
'80s \ce{Be2} represented a technical challenge for the experimentalists, and 
even later the value of its binding energy was not well established.
A review on the experimental and theoretical investigations of \ce{Be2}
has recently appeared\cite{be2science},
containing also new reference experimental data for its binding energy.

\begin{figure}
\centering
\includegraphics[width=1.\columnwidth]{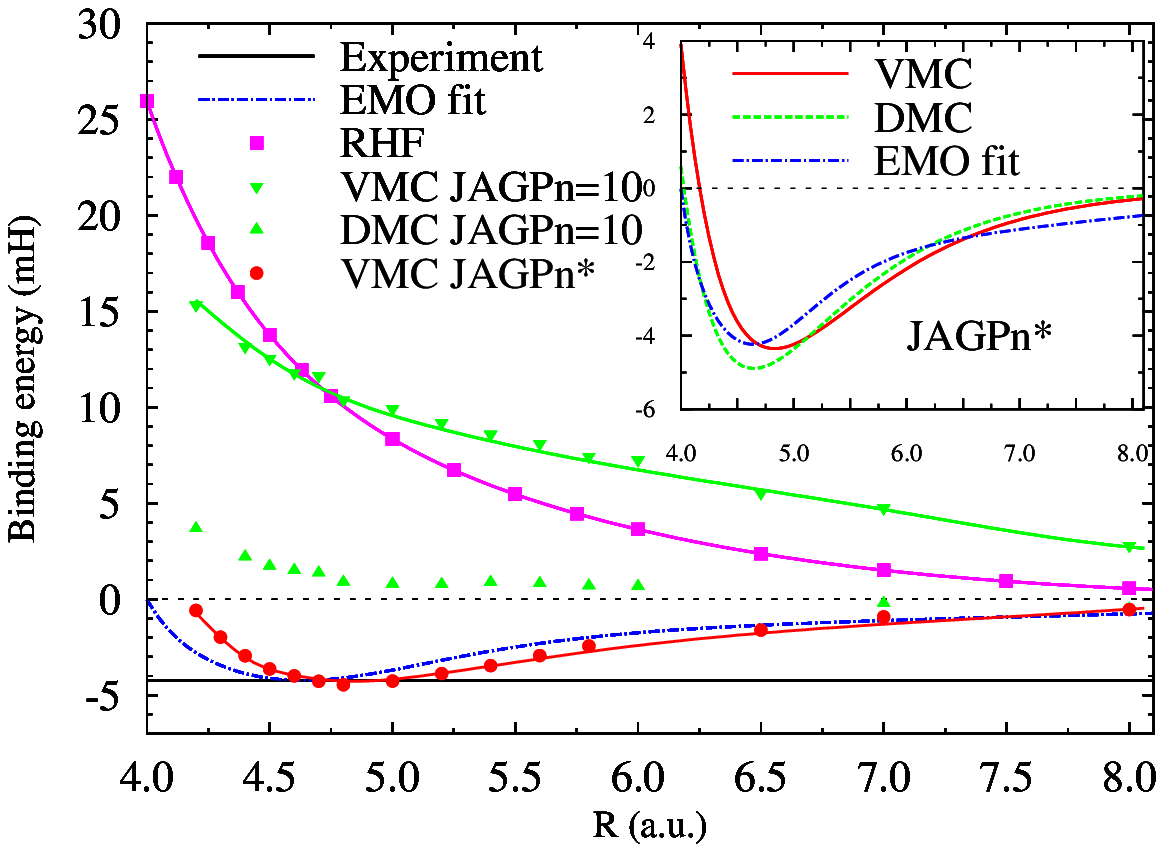}
\caption{\small{\ce{Be2} binding energy (in milli-Hartree) versus the nuclei 
distance $R$ (in 
atomic units). Comparison of RHF\cite{milestone} outcomes, 
JAGPn$^*$ results (in this case $n^*=4$), and
VMC and LRDMC results obtained with a JAGP wave function with $n=10$ 
(squares, dots, downward and upward
triangles respectively, while lines are a guide to the eye). 
In the figure, 
the experimental binding energy subtracted 
by the zero point energy\cite{be2science} (solid line), and an EMO fit of the 
experimental data\cite{be2science} (slash-dotted line) are also plotted.
The reference atomization limit for the JAGPn$^*$ results is
given by atomic calculations with a JHF wave function. For $n=10$, the
atomization reference is given by an atomic JAGP wave function with the same
primitive basis set as the JAGP wave function for the dimer.
In the inset: comparison of fits to VMC and LRDMC data 
(solid and dashed lines respectively) for the JAGPn* wave function and the 
EMO fit of experimental data from Ref.~\cite{be2science} (slash-dotted line). 
Labels for the inset axes are the same as in the main frame.
All curves are shifted with respect to their own 
atomization limit.}}
\label{be2final}
\end{figure}

In Fig.~\ref{be2final} we provide the energy dispersion curve for the \ce{Be2} 
molecule. 
The main plot shows a comparison between standard RHF 
calculations\cite{milestone}, VMC data obtained with the JAGPn$^*$ wave 
function, VMC and LRDMC  results for a JAGP with $n> n^*$.
We also show an expanded Morse oscillator (EMO) fit of the recent experimental 
data of Ref.~\cite{be2science}. 

As mentioned before, in this case it turns out that 
$n^* \le n_{HF}(\ce{Be2})+1$.
In particular, by using the upper bound of Eq.~(\ref{ntilde})
our JAGPn$^*$ reduces to a 
simple JHF wave function with $n^*=4$ [the upper bound $n^*=5$ 
of Eq.~(\ref{therightn}) does not provide 
significant improvements in  a fully symmetric calculation].
Within the  $n=n^*$ constraint,
bond features such as binding energy
and bond length are reproduced fairly accurately, 
whereas a trial wave function with $n>  n_{HF}(\ce{Be2})+1$ fails to bind the 
molecule at the expected distance, 
even though the total VMC (LRDMC) energy $ E =-29.32295(8)$H ($E=-29.33385(7)$H) is much below the constrained 
minimization by about $24$mH ($14$mH) at $R=5$ a.u..
This total energy is very accurate from an absolute point of view and 
compares well with state of the art QMC calculations.\cite{umrigar}
However the variational wave function with the 
lowest variational energy, i.e. the JAGP with 
$n=10$, behaves similarly to an uncorrelated RHF, 
and both provide a very poor description of this chemical bond.\cite{be2limit} 
 More in detail, 
the VMC JAGP$n=10$ energy dispersion curve presents one minimum at 
an interatomic distance $R>8$ a.u., while LRDMC JAGP$n=10$ displays
an additional swallower minimum close to the expected bond length.
On the other hand, the quite accurate dispersion curve obtained by 
the full optimization of the JHF wave function shows,  
for the first time to our knowledge, that the key missing ingredients in the 
HF for \ce{Be2}  are  just the dynamical correlations carried out by 
our Jastrow factor.
Though very simple, our Jastrow factor includes many-body correlations 
(up to two-ion two-electron interactions), that allow to take into account 
effective attractions between atoms given by vdW forces,\cite{rocca}
and other polarization-polarization contributions.\cite{beaudet} 
Indeed, the dynamical interactions are extremely important to bind the 
molecule and it is crucial that the Jastrow factor includes this effect. 
For instance, the different parametrization of the Jastrow factor 
used in Ref.~\cite{umrigar} does not allow to bind \ce{Be2} 
at a variational level, at variance with this work. 
On the other hand, the DMC binding energies of Ref.~\cite{umrigar}
are much closer to our VMC and DMC results, 
further suggesting  the importance of the dynamical correlations in the bond.

In the inset of Fig.~\ref{be2final}, we compare the VMC and LRDMC JAGPn$^*$ 
energy dispersion curves shifted by their asymptotic limits. 
Despite some slight differences,
the agreement between the two QMC techniques within the $n=n^*$ constraint 
and the most recent experimental findings\cite{be2science}
can be considered fairly good in this case, due to
the very weak binding of the molecule.

Also the JAGPn$^*$ description for \ce{B2} reduces to a JHF wave function.
Both bond length and binding energy agree within two standard deviations with
the estimated exact data.

\subsection{Fluorine dimer}
\begin{figure}
\centering
\includegraphics[width=0.7\columnwidth]{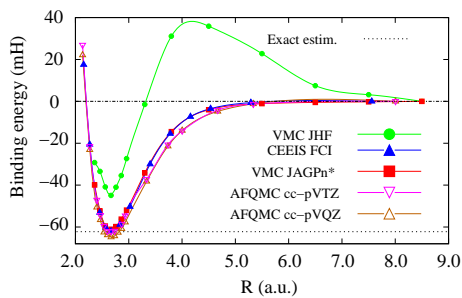}
\caption{\small{Energy dispersion curve for the \ce{F2} molecule 
obtained with several computational techniques. 
The zero reference energy (dashed line) is 
twice the JHF atomic total energy for the JAGPn$^*$ and the JHF molecules, 
whereas
it is the large distance energy of the dimer for the remaining 
data. 
The slash-dotted line indicates 
the experimental binding energy\cite{bytautas} subtracted by the zero point and
spin orbit energies. Lines are a guide to the eye. 
The CEEIS-FCI data are taken from Ref.~\cite{f2fci_corrections}, whereas AFQMC
data are taken from Ref.~\cite{f2_shiwei}.}}
\label{fig:F2}
\end{figure}

A remarkable example of the accuracy of our technique is provided by the
energy dispersion curve of the fluorine dimer reported in Fig.~\ref{fig:F2},
where we show the results obtained with various QMC methods 
(and different wave functions), and other ab-initio results.
More in detail, we compare our JHF and JAGPn$^*$ VMC data (see also 
Table~\ref{tabf2}, where, for comparison, we also report our LRDMC results)
with two
energy dispersion curves obtained with auxiliary-field QMC (AFQMC) simulations
for an unrestricted HF reference wave function 
spin-projected 
to eliminate spin contamination
\cite{f2_shiwei}, and an ab-initio study based on full 
configuration interaction (FCI) calculations combined with the 
correlation energy extrapolation by intrinsic scaling (CEEIS) 
technique\cite{f2fci} plus core-electron correlations and scalar relativistic 
corrections\cite{f2fci_corrections}. 
\begin{table*}
\begin{center}
\begin{threeparttable}
\caption{\small{\ce{F2} binding energies (in mH) as a function of the 
internuclear distance $R$ in a.u. (see also Fig.~\ref{fig:F2}): VMC and LRDMC 
results (energy of the molecule at distance $R$ minus two times the JHF atomic 
energy) for a JHF and the JAGPn$^*$ wave functions.}}
\label{tabf2}
\begin{tabular}{|| c| c| c||c|c|| }
\hline
 R (a.u.) & VMC JHF & VMC JAGPn$^*$ &  LRDMC JHF &LRDMC JAGPn$^*$  \\
\hline

2.36  & -23.9(4) & -39.9(4)& -32.4(4) &-39.6(4)\\         
2.46  & -28.1(6) & -52.3(5)& -42.1(5) &-53.2(4)\\         
2.56  & -35.5(5) & -59.5(5)& -48.2(5) &-59.7(4)\\         
2.668 & -39.6(5) & -61.0(4)& -49.7(4) &-60.9(4)\\         
2.76  & -35.6(5) & -59.7(5)& -49.1(4) &-60.6(4)\\         
2.86  & -30.1(4) & -56.2(4)& -44.1(4) &-57.1(4)\\         
2.96  & -22.0(5) & -51.9(4)& -38.1(5) &-52.1(4)\\         
3.3   &   3.9(5) & -34.2(5)&          &-34.6(4)\\         
3.8   &  36.5(6) & -14.4(5)&          &-13.4(4)\\         
4.5   &  41.2(6) & -4.0(4) &          &-2.5(4)\\         
5.5   &  28.1(5) & -1.0(4) &          &       \\         
6.5   &  12.8(5) & -0.1(5) &          &       \\         
7.5   &   8.5(5) &  0.5(5) &          &       \\         
8.5   &   5.4(5) &  0.2(5) &          &       \\

\hline

\end{tabular}
\end{threeparttable}
\end{center}
\end{table*}
   
One can observe the dramatic improvements of the JAGPn$^*$ wave function
with respect to JHF simulations (see also 
Table~\ref{tab_molecules}--\ref{tabf2}). 
According to Eq.~(\ref{ntilde}), we have used $n^*=n_{HF}(\ce{F2})+1$,
because the upper bound of Eq.~(\ref{therightn}), $n^*=n_{HF}(\ce{F2})+2$, does not lead 
to significant differences within our energy accuracy.  
We remark here instead the importance of adding just one 
molecular orbital to the Hartree-Fock theory, because this allows to 
consider all bonding and antibonding MO's 
in the AGP, thus leading to a fully size consistent result which benchmarks 
the energy dispersion curve from the bond length to the atomization limit.
The agreement of the JAGPn$^*$ with the ab-initio CEEIS-FCI calculations is remarkably 
good already at a VMC level. In fact, the VMC binding is 1.671(2) eV against 
1.6867 eV of FCI calculations (without spin-orbit
corrections). The LRDMC binding is 1.688(2) eV. 
Instead the AFQMC curves seems to be shifted of 
approximately 2-3mH with respect to our JAGPn$^*$ data in the bond and 
intermediate length regions. This is due to an underestimation of the energy
at large distance caused by the use of a simple unrestricted HF wave function
(see the discussion in Section IV of Ref.~\cite{f2_shiwei} and Figs. 4 and 6
therein). Indeed the AFQMC well depth is 1.70(2) eV and 1.77(1) eV for the 
cc-pVTZ and  the cc-pVQZ wave functions respectively, when the reference at 
large distance is the molecular energy, whereas it is 1.60(1) eV and 1.70(1) eV 
respectively, when the large distance reference is twice the 
energy of the separated atoms.

The results we have presented so far,
reported in Table~\ref{tab_molecules},
represent an astonishing example of 
the importance of constraining the 
variational wave function to an appropriate form 
during the optimization of the energy.
Indeed, a brute force optimization of 
a correlated wave function, which is 
a rather demanding computation  especially within 
QMC,  would lead to an upper bound of the total energy which is almost 
meaningless, particularly in the \ce{Be2} case.
The rational behind this effect is that an unconstrained optimization 
may not satisfy the requirement for the wave function to be a fair 
representation of the ground state of a physical Hamiltonian.
While in lattice models it is possible to constrain the determinantal 
part of the RVB wave function
to be the ground state of a short-range Bardeen-Cooper-Schrieffer (BCS) 
Hamiltonian --a quite sensible and accepted choice in strongly 
correlated models-- this is much harder in continuous-type calculations.
The constraint that we propose, 
very simple to implement in practice, just mimics the effect 
of computing the ground state of an HF Hamiltonian with an additional sufficiently 
weak BCS coupling  between electrons. 
In fact, in this limit one obtains  the 
complete or partial occupation --via the $\lambda_j$ in Eq.~(\ref{pairing})-- 
of a number  of  molecular orbitals $n^*$ not necessarily equal to the 
RHF prediction.
In this context, the BCS coupling represents the effective interaction
between electrons, which pairs them into the chemical bond.
For instance, it is  well known that  
the ground state of the \ce{H2} dimer at large distance 
is very well described by  
the singlet entangled state obtained with the AGP\cite{barbiellini},
only when the bonding and antibonding orbitals are taken into account.
This state can be considered the ground state of a BCS Hamiltonian 
that in the atomization limit simply splits  into a sum of two atomic 
HF Hamiltonians, with vanishingly small pairing.  This coupling 
 is however  important to lift the degeneracy  between the singlet and  
the triplet states. 
The same physics happens in the \ce{F2} molecule studied in this work.

\subsection{Carbon, nitrogen, and oxygen dimers} 
The  \ce{C2}, \ce{N2} and  \ce{O2}  molecules represent challenging cases
for our correlation consistent AGP approach.
Indeed, when $0\le S<S_{max}$ and $S_{max}>1$, 
the lack of size-consistency in the AGP poses a 
fundamental limitation in order to reach the JHF limit in the dissociation.
Note that $S=0$ and $S_{max}=1$ is a non trivial case when
the JAGP is size consistent and the JHF is not 
(e.g. the simplest \ce{H2} molecule, or the \ce{F2} described in the previous 
section). 
Strictly speaking, the restriction of the number of MOs  
to $n^*$ does not guarantee a size consistent JHF result,
even for the JAGP wave function. In the general case, 
the total JAGP energy in the atomization limit 
is an upper bound of $ \sum_A E_{JHF}(A) $ evaluated in the CBS limit.
A generalization of the JAGP, based on the pfaffian algebra, which includes 
also triplet pairing 
for electrons with the same spin, allows to have JHF size consistent results 
also in the cases with $S=S_{max}-1$ ($S_{max}>1$), 
e.g. in \ce{O2}.
Despite we have not implemented this generalization,  
triplet pairing seems to provide a rather negligible effect at bond length, 
as very good 
results can already be obtained with the present JAGP ansatz. 

The constrained minimization of the JAGPn$^*$ wave function leads to very significant 
improvements with respect to JHF results in both the binding energy and the bond length for \ce{C2} and \ce{N2},
if compared with the exact estimates coming from the experimental values.
In \ce{O2} the situation is quite different, since
$n^*=n_{HF}(\ce{O2}) +1$
and the molecular orbital missing in the HF scheme is quite high in energy, 
compared to all the paired  molecular orbitals included in 
Eq.~(\ref{pairing}) according to our constraint.
Therefore, it is not surprising that, in this case, 
the improvement upon JHF findings is smaller.
The comparison with the exact estimates is nevertheless quite satisfactory.

By comparing the results in Table~\ref{tab_molecules}
for \ce{N2}, \ce{O2} and \ce{F2} with Ref.~\cite{umrigar}'s
single-determinant DMC well depths, we note improvements
already at the JHF level, and even
bigger improvements are obtained  upon  Grossman's benchmarks\cite{grossman} (to which 
one should add the ZPE).
We should mention that in Ref.~\cite{grossman}, the basis used for the
determinantal part is not optimized. Other differences could be due 
to either the different pseudopotentials used or the calculations not fully converged in 
the CBS limit.

\begin{table}
\begin{center}
\begin{threeparttable}
\caption{\small{
Well depths for \ce{C2}, \ce{N2}, \ce{O2}: comparison between 
pfaffian results\cite{pfaff1,pfaff2} and JAGPn$^*$ results.
The VMC and LRDMC findings are compared with the exact estimates previously
reported in Table~\ref{tab_molecules}.
}}
\label{tab_pfaff_jagp}
\begin{tabular}{|l l|  c| c | c| }
\hline\hline
 Method & WF& \ce{C2} & \ce{N2} & \ce{O2}\\
\hline
VMC   & STU       & 5.94(2)  & 9.42(3)  & 4.94(3)\\
VMC   & JAGPn$^*$ & 6.327(9) & 9.874(2) & 5.060(7)\\
DMC   & STU       & 6.26(2)  & 9.84(2)  & 4.93(2)\\
LRDMC & JAGPn$^*$ & 6.297(8) & 9.882(1) & 5.125(5)\\
\hline
\multicolumn{2}{|c|}{Exact.  estim.} &  6.44(2)& 9.908(3)& 5.241(3)\\
\hline\hline
\end{tabular}
\end{threeparttable}
\end{center}
\end{table}
As a further evidence of the accuracy of our method, 
in Table~\ref{tab_pfaff_jagp} 
we compare the well depths found for \ce{C2}, 
\ce{N2} and \ce{O2} with results obtained
by using a singlet-triplet-unpaired (STU) pfaffian wave 
function\cite{pfaff1,pfaff2} 
and taking as a reference at large distance the JHF atomic limit,
i.e. the binding energy is computed as $E_{molecule}(\hbox{STU})-2\,E_{atom}(\hbox{JHF})$.
As shown in Table~\ref{tab_pfaff_jagp}, we find a rather good agreement for the 
well depth even in the challenging \ce{C2} molecule, 
when $S= S_{max}-2$ and an exact size consistent result is not 
possible even within the more general pairing function containing triplet 
correlations.
In principle, the pfaffian wave function, being a generalization of the JAGP, should
have a larger binding energy by taking the corresponding 
JHF atomic energy as a reference, provided 
the CBS limit is reached and the pseudopotentials are accurate enough.
Instead, in all the cases shown in Table~\ref{tab_pfaff_jagp},  
the STU and the JAGPn$^*$  binding energy are very close, 
at least at the DMC level.
The fact that our binding energy is always larger and more accurate
comes  probably from the use of a more complete basis, 
fully optimized in both the coefficients and the exponents, whereas in 
 Ref.\cite{pfaff1,pfaff2} the atomic basis is not optimized.

These comparisons show that our JAGPn$^*$  ansatz 
provides quite generally a very accurate description of the chemical bond.
This emphasizes the role of singlet electron pairs in the chemical bond,
and is consistent with the RVB theory. 
Thus, the present variational wave function can be considered 
the cheap but nevertheless accurate realization of the RVB idea,
since only a single determinant and standard variational Monte Carlo are needed.

\subsection{Heteronuclear dimers}
We further carried out calculations for a couple of heteronuclear diatomic molecules
belonging to the G1 set (namely \ce{LiF} and \ce{CN}), selected on the basis of 
the quite big discrepancy between the binding energy found in
Ref.~\cite{grossman} and the reported experimental values.
Bond lengths, dissociation energies and ZPEs for these two molecules are 
reported in Table~\ref{tab_molecules}.
We compare our well depths with the exact estimates obtained by correcting 
the experimental dissociation energies with 
the experimental ZPE and spin orbit energies reported in 
Ref.~\cite{feller}.

As mentioned in Section \ref{molnumber}, \ce{LiF} is one of the cases in which
$n^*=\tilde n$ does not yield significant improvements with respect to the
JHF wave function, even though $\tilde n>n_{HF}$.
Instead, for \ce{CN} the JAGPn$^*$ wave function improves the description of the
bond with respect to the JHF one, giving a bond length in fairly good agreement
with the experimental value, although the binding energy is underestimated by
$\approx 0.1$ eV with respect to the exact estimate reported in 
Ref.~\cite{feller}. 
As for the homonuclear dimers shown in the previous sections, also in the 
heteronuclear cases here considered, our method provides
binding energies in closer agreement with the experimental 
values\cite{feller} than those of  Ref.~\cite{grossman}.
In particular, for \ce{LiF} the agreement is very good 
already at the JHF level, as anticipated above. 

\section{Conclusions}

In conventional quantum Monte Carlo 
variational techniques, based on the use of the Jastrow factor, it is not 
possible to consider a finite basis set and to exploit the huge cancellation 
between atomic energies and molecular energies within the same basis set.  
Indeed, after  the introduction 
of the Jastrow factor, the wave function is unavoidably defined on an infinite 
dimensional Hilbert space. As a consequence, it is more difficult 
to achieve  the chemical accuracy on
the energy differences and obtain  a good description of the chemical 
bond, as we have shown for instance in the \ce{Be2} case.
Here,  a very accurate variational energy obtained by applying the DMC 
technique to our lowest energy JAGP wave function,  
completely misses the correct features of the bond. In this case, with 
an unconstrained variational approach,  
qualitatively correct results can be obtained
probably only by reaching the 
chemical accuracy on the total energy, that is clearly
a very difficult task for any approximate variational technique.
In fact, this target was so far achieved within QMC only by
using several determinants in small molecules\cite{umrigar}.
 
In this paper we propose a simple constraint which allows to exploit
the above mentioned cancellation between atomic and molecular energies 
even in QMC calculations based on a single determinant 
wave function. In fact, 
instead of imposing a constraint on the dimension of the atomic 
Hilbert space we 
change a bit this  
point of view by constraining the number of molecular orbitals 
to an appropriate value that allows to take JHF results for the isolated 
fragments as a reference for the dissociation energy of the molecule.
With this constraint we have shown  that it is possible to obtain much 
more accurate results in both variational and LRDMC calculations.

Although we have not carried out a systematic study of all the G1 set considered in 
Ref.~\onlinecite{grossman}, in several cases where the discrepancy was 
sizeable we obtain an almost exact description of the bond (e.g. in \ce{F2}).
Surprisingly the LRDMC calculation provides only small improvements upon 
the simple and much cheaper VMC calculation, 
which turns out to be remarkably accurate in our approach.
Also in cases where we do not improve upon the JHF results (e.g. \ce{LiF}), 
we nonetheless obtain 
accurate binding energies, within a precision of about 0.1 eV.
The latter achievement could be due to 
the accurate basis set  we have considered in our work,
together with the state-of-the-art optimization technique,\cite{cyrc2} 
which is able to handle the large number of parameters in an extended basis set.

In conclusion, we have introduced a new and general approach to perform 
electronic structure calculations of quantum chemistry compounds
based on a variational RVB wave function.
In this formulation, we have shown that a substantial improvement
in the description of the chemical bond is possible by 
extending the  standard correlated single determinant theory with the JAGP 
wave function.
In the original formulation of the RVB theory, the gain in energy obtained by 
the resonance of several valence bond configurations was just named the 
'resonance valence bond energy'. 
Within this new formulation we propose that this energy 
gain can be achieved 
by  increasing  the number of molecular orbitals of the JAGP 
from its HF value, and without exceeding a value $n^*$ of molecular orbitals. 
This value  can be determined by requiring  
a correlation consistent property
from the bond length to the atomization limit, realized via a constrained 
energy minimization.

\acknowledgements

This work was partially supported by COFIN2007, and CNR. 
One of us (M.C.) acknowledges 
the Centre de Physique Th\'eorique of the
Ecole Polytechnique.
We thank Shiwei Zhang for useful comments.

\appendix
\section{ Constrained optimization of the AGP wave function }
\label{molorb}
\subsection{Molecular orbital expansion of the AGP}
In this appendix 
we expand the 
pairing function $\Phi$  in atomic orbitals 
$\xi_{j}( \vec r)$  located at atomic 
positions $\vec R_j$:
\begin{equation} \label{wf} 
\Phi(\vec r, \vec r^\prime)=\sum_{j,j^\prime} \lambda_{j,j^\prime}
\xi_{j} (\vec r) \xi_{j^\prime} (\vec r^\prime), 
\end{equation}
where $\lambda$ is the pairing matrix 
and 
$j,j^\prime$ label  the considered atomic orbitals  on the 
corresponding atomic positions $\vec R_j , \vec R_{j^\prime}$.
Obviously, in order to define a singlet state 
the pairing matrix should be 
symmetric $\lambda_{j,j^\prime} = \lambda_{j^\prime,j}$.
Hereafter, both for simplicity and for the sake of 
 generality we do not assume this symmetry, because  
it can be easily satisfied during the optimization scheme, when necessary.
Therefore, in the general case 
we are left with $N_L=L \times L-1$ independent variational constants, 
where $L$ is the linear 
size of the matrix $\lambda$, namely the dimension of the atomic basis.
There is only one linear dependence between the $L \times L$ entries  of the matrix 
$\lambda$ because the multiplication of $\Phi$ by an overall constant 
does not change the AGP apart for its normalization.
This constraint is usually satisfied by keeping fixed an arbitrary matrix 
element to the unit value.
 
Usually, the number $N_L$ is very large and in the following we determine 
a systematic way to work with much less variational parameters,
being nevertheless efficient in determining the lowest energy molecular 
orbitals of the chosen variational ansatz.

For simplicity we do not consider 
unpaired orbitals, because for them no constraint is applied, 
therefore we set  $N_\uparrow=N_\downarrow$.
Moreover, in the following we can assume that 
the original 
orbitals $\xi_j$ have been orthogonalized by a suitable transformation
depending on the overlap matrix
\begin{equation} \label{overlap} 
S_{i,j} = \langle \xi_i| \xi_j \rangle, 
\end{equation}
namely we implicitly assume the following change of the definition 
of the orbitals and the corresponding matrix $\lambda$ in Eq.~(\ref{wf}):
\begin{eqnarray} \label{transformm}
 \xi_j (\vec r) &\to & \sum\limits_k (S^{-1/2})_{j,k} \xi_k (\vec r),  \nonumber \\
 \lambda_{i,j} &\to & \sum\limits_{k,l} (S^{1/2})_{i,k} \lambda_{k,l}  (S^{1/2})_{l,j}. 
\end{eqnarray}
This greatly simplifies the forthcoming analysis without loss of generality.

Then, for the resulting 
  square matrix we can use the well known singular value decomposition:
\begin{equation} \label{basic}
\lambda_{i,j} = \sum\limits_{k=1}^r  \alpha_k  \psi^k_i \bar \psi^k_j,
\end{equation}
where $\alpha_k\ge 0$ and $\psi^k$ ($\bar \psi^k$)  are a set of $r \le L$ 
molecular orbitals for the spin-up (spin-down) electrons 
that are orthonormal, i.e. $\sum_l \psi^i_l \psi^j_l= \delta_{i,j}$.
Formally the spin-up molecular orbitals and the spin-down ones 
are the eigenvectors of the 
$2L \times 2L$ symmetric matrix 
\begin{equation} \label{maptosym}
H= \begin{array}{|cc|}
0 & \lambda \\
\lambda^{\dag} & 0 
\end{array}
\end{equation}
which has pair of eigenvectors with eigenvalues 
 $\pm \sqrt{\alpha_k} $ given by:
\begin{equation}
\begin{array}{|c|}
 \psi^k \\
\pm \bar \psi^k 
\end{array}\,.
\end{equation}

A simple way to reduce the number of parameters is to require that the 
matrix has rank $r < L$ so that all the eigenvalues   $\alpha_k $ 
for $k>r$ are assumed to be zero or negligible. 
For instance if $r=N/2$ we obtain the standard Slater determinant 
with $N_\uparrow=N_\downarrow=N/2$ molecular orbitals for each spin component.

This projection scheme can be made general, and this leads to a remarkable 
extension  of the Slater determinant, within the AGP wavefunction 
expanded in molecular orbitals, as discussed in the forthcoming subsection.

\subsection{ Projection on a rank-$r$ geminal}
If the rank $r$ of a geminal matrix $\lambda_{i,j}$ is equal to half 
the number of electrons $N/2$, then the AGP  represents a   
Slater determinant. Even if $N/2$ is usually much smaller than the dimension 
of the atomic basis $L$,  Fermi statistics at zero temperature favors the 
occupation of the lowest possible energy levels, so that 
$r \simeq N/2$  turns out to be a  
reasonably accurate guess for the AGP wave function. In principle 
this wave function  may have 
much larger rank up to $r=L$, but one may expect that 
 most of the singular values will have negligible 
weight.
Therefore, from a general point of view, and 
not only for reducing the number 
of variational parameters, it is important to optimize in an 
efficient way a full  $L \times L$ matrix of rank-$N/2 \le r \ll L$ given by 
Eq.~(\ref{basic}).

To this purpose we propose the following scheme of constrained optimization, 
where $r$ is chosen and fixed to a reasonable value $n^* \approx N/2$
during the optimization.

Given $\lambda^0$ a rank-$r$ matrix, in order to  
simplify the notations,  
we write 
the corresponding singular value decomposition (\ref{basic}) in a matrix form:
\begin{equation}
\lambda^0 = \psi^0 \alpha^0 \bar \psi^0_T,
\end{equation}
where 
$\psi^0$ and $\bar \psi^0$ 
are $L\times r$ matrices,  
the subscript $T$ indicates the transpose of a  matrix,   
and the non zero singular values
$\alpha^0_k$, $k=1,\cdots r$ are denoted by a diagonal matrix $\alpha^0$. 

Then  we change 
this  matrix $\lambda^0$ by adding to it  a general first order contribution:
$$ \lambda^\prime= \lambda^0 + \lambda^1 (\epsilon),$$
where henceforth the superscript indicates the order of the expansion in 
$\epsilon$. 
This new matrix will be constrained to have rank $r$.
Therefore all the terms in Eq.(\ref{basic}) can be expanded within first 
order in perturbation:
\begin{eqnarray} \label{project}
 \lambda^1  &= & \epsilon  (\psi^0  \alpha^0 \bar \psi^{1}_T +  \psi^1  \alpha^0 \bar \psi^{0}_T )   \nonumber\\
&+& \epsilon  \psi^0 \alpha^{1} \bar \psi^{0}_T.
\end{eqnarray}

In order to satisfy the constraint on the rank in the matrix $\lambda^1$, 
it is much simpler to work with an unconstrained matrix $\bar \lambda$, 
and left and right projection matrices:
\begin{eqnarray} \label{matrixform}
P^R &=& \bar \psi^0_T \bar \psi^0, \\ 
P^L &=&  \psi^0_T  \psi^0. 
\end{eqnarray}
The two matrices above are projection matrices ($ P = P_T$ and $P^2=P$) 
as they project vectors in the $r$  dimensional subspaces 
corresponding to the non zero values of the singular value decomposition 
(\ref{basic}).

Indeed it is very simple to show that if the matrix $\lambda^0+\lambda^1$ 
satisfies 
 the constraint of a singular value decomposition with rank $r$, 
$\lambda^1$ has to satisfy the simple relation:
\begin{equation} \label{rcondition}
 (I -P^L) \lambda^1 (I -P^R) =0, 
\end{equation}
because in the expression (\ref{project})  $ (I-P^L) \psi^0 =0 $ and 
$  \bar \psi^0_T (I -P^R) =0$. 

Thus an unconstrained variation of the matrix $\bar \lambda$ 
can be projected onto  
the constrained one  by using the above projection matrices:
\begin{equation} \label{finaleq}
\lambda^1  =  \bar \lambda - (I -P^L) \bar \lambda ( I -P^R) 
\end{equation}
in the sense that, after the above projection, the matrix $\lambda^0 +  
\lambda^1 $ is  suitable 
and can be considered to satisfy the constraint of a rank-$r$ matrix 
at first order in the perturbation (the matrix
 $\lambda^1$ being sufficiently small).

Indeed, by simple inspection, the RHS  of Eq.~(\ref{finaleq}) immediately 
satisfies the condition (\ref{rcondition}), that is so far considered a 
 necessary condition.
It is also possible to show with a lengthy but straightforward 
calculation using first order perturbation
theory of the  symmetric matrix $H$ given in (\ref{maptosym}), that relation
(\ref{rcondition}) is also a sufficient condition for a perturbation 
that does not change the rank of a singular value decomposition.
 
\subsection{Application to QMC} 
In the actual application of the recent QMC scheme for minimization of the 
energy, it is important to evaluate derivatives of a function with respect 
to the  unconstrained parameters $\bar \lambda$.
This function $E(x)$ can be  either the logarithm of the 
wave function or the local energy on a particular electronic 
configuration $x$ sampled by the MC technique.\cite{umrigar,casula1,casulamol}.

Given the matrix $ D_{i,j}=\partial E / \partial \lambda_{i,j}$
of the unconstrained derivatives with respect to $ \lambda_{i,j} $, 
by using Eq.~(\ref{finaleq}) 
and  the chain rule for derivatives, then the corresponding matrix of 
constrained derivatives $ \bar D_{i,j}=\partial E / \partial \bar \lambda_{i,j}$
can be computed by simple  matrix manipulation in the following way:

\begin{equation} \label{finald}
 \bar  D =  D - (I-P^L_T) D (I-P^R_T). 
\end{equation}

In order to work with the original matrices we have to replace in
 Eqs.~(\ref{finaleq},\ref{finald}) the ones obtained by applying the 
inverse of the transformation (\ref{transformm}):
\begin{eqnarray}
P^R &\to & S^{1/2} P^R S^{-1/2},  \\
P^L & \to & S^{-1/2} P^L S^{1/2}.
\end{eqnarray}
Notice also that after  this transformation $P^R_T$ and $P^L_T$ are no longer 
equal to $P^R$ and $P^L$ in Eq.~(\ref{finald}).

The scheme therefore can be summarized in the following steps:
\begin{enumerate}
\item Compute the unconstrained derivatives $D$ that, with some algebra, 
can be casted into  a product of much smaller rectangular matrices $U,V$ 
such that $D= U_T V$ of dimension $L \times N/2$. 
Notice that also the projection matrices can be written in this 
convenient form, as in Eq.~(\ref{matrixform}). 
\item   
Apply the projection (\ref{finald}) by using the current molecular 
orbitals. By exploiting the fact that all 
the matrices involved  are written in terms of much smaller rectangular 
matrices, a very convenient computation can be achieved 
scaling like $N^2 L$ instead of 
$L^3$ as in the straightforward implementation of the projection.  
\item Apply the recent optimization schemes\cite{umrigar,casula1,casulamol},
and change the unconstrained parameters $\bar \lambda$.
Then apply the projection, by diagonalizing the matrix $\bar \lambda$ and taking only 
the right and left eigenvectors corresponding to the 
largest  singular values.  New molecular orbitals are then defined after 
this diagonalization. 
\item Repeat the above-described steps until convergence in the energy  is 
achieved.
\end{enumerate}

\section{AGP and basis set expansion effects}
\label{check_app}
 
\subsection{Effect of the improved upper bound for $n^*$: the \ce{C2} case}
\label{c2molapp}
\begin{figure}
\centering
\includegraphics[width=1.\columnwidth]{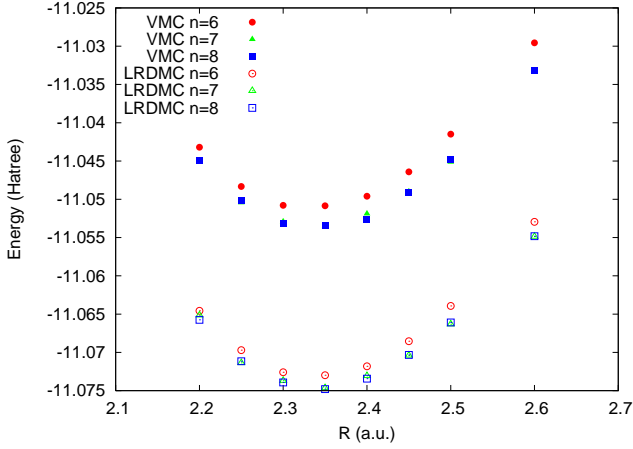}
\caption{\small{Energy dispersion curve for \ce{C2} dimer with
$n=6$ (circles), $n=7$ (squares), $n=8$ (triangles) molecular 
orbitals. VMC and LRDMC results are represented by filled and empty 
symbols respectively.}}
\label{c2mol}
\end{figure}

As explained in Section \ref{molnumber}, \ce{C2} is one of
the exceptions to the rule of Eq.~(\ref{ntilde}). In this case,
we have $\tilde n=6$, but the more accurate 
upper bound in Eq.~(\ref{therightn}) allows 
to work with $n^*=7$. 
Indeed, by following strictly Eq.~(\ref{ntilde}) 
one would include a single antibonding orbital $\pi^*$ in the AGP, 
while that orbital is double degenerate, due to the rotational symmetry of the molecule. 
Therefore, in order to fulfill the symmetry 
of the dimer, it is particularly important to fill 
 the degenerate levels in the AGP  
by setting $n^*=7$. 
In Fig.~\ref{c2mol} we show the VMC and LRDMC 
energies at various internuclear distances 
found with a JAGP wave function expanded 
in $\tilde n$ and $n^*$ molecular orbitals. 
In this case, the 
improved upper bound for $n^*$ yields a gain of 
$\approx 2.7$ mH in the VMC energies and of 
$\approx 1.5$ mH in the LRDMC ones.
Incidentally, the $n=n^*=7$ energies agree within the error bars
with the data resulting from a JAGP with $n=8$ molecular 
orbitals.

An analogous check was done with all electron simulations at a fixed
interatomic distance $R=2.35$ a.u..
Results are reported in Table~\ref{allelc2nmol}.
We note a saturation of LRDMC total energies for $n>n^*$. 
\begin{table}
\begin{center}
\caption{\small{Total energies from all electron calculations for \ce{C2}
with a JAGP expanded on a different number of MOs $n$ are shown. 
$n=9$ corresponds to the $n^*$ upper bound of Eq.~(\ref{therightn}).
The estimated exact total energy $E_0$ is reported for comparison.}}
\label{allelc2nmol}
\begin{threeparttable}

\begin{tabular}{|ccc|} 
\hline\hline 
    n   & VMC & LRDMC \\
\hline
9      &  -75.8439(7)  &   -75.8934(4)\\
10     &  -75.8453(7)  &   -75.8930(4)\\
15     &  -75.8473(8)  &   -75.8928(4)\\
\hline
$ E_0$ & \multicolumn{2}{c|}{-75.9265\tnote{a}} \\
\hline\hline
\end{tabular}

\begin{tablenotes}
{\footnotesize
\item[a]{From Ref.~\cite{bytautas}}
}
\end{tablenotes}

\end{threeparttable}
\end{center}
\end{table}

\subsection{Convergence in the basis set of the AGP and the Jastrow parts}
\label{basisconvergence}
Below we report the convergence in the basis set for selected molecules.

\subsubsection{\ce{N2}  convergence in the basis set}

For \ce{N2} we checked the convergence in the basis set by means of VMC 
and LRDMC simulations at the experimental internuclear distance.
Total energies are reported in Table~\ref{n2_total_energies}.
\begin{table}
\begin{center}
\caption{\small{JHF and JAGPn$^*$ VMC and LRDMC total energies (in 
Hartree) for the \ce{N2} pseudo-molecule. 
Energies were computed at the experimental bond length using
different basis sets for both the determinantal part 
and the three and four-body Jastrow term (34BJ).}}
\label{n2_total_energies}
\begin{tabular}{|ll|cc|} 
\hline\hline
\multicolumn{4}{|c|}{JHF}\\ 
\hline
    Det.       & 34BJ         &      VMC       &    LRDMC\\
\hline
$5s5p$   &  $2s2p$   &  -19.9031(5)   & -19.9338(2)\\
$5s5p$   &  $3s3p$   &  -19.9071(3)   & -19.9346(2)\\
$5s5p$   &  $4s4p$   &  -19.9076(4)   & -19.9349(2)\\
$4s4p$   &  $3s3p$   &  -19.9071(3)   & -19.9344(2)\\
$6s6p$   &  $3s3p$   &  -19.9080(3)   & -19.9348(2)\\
$5s3p2d$ &  $2s2p$   &  -19.9074(3)   & -19.9339(1)\\
$5s5p2d$ &  $3s3p$   &  -19.9086(4)   & -19.9349(2)\\ 
\hline\hline
\multicolumn{4}{|c|}{JAGPn$^*$}\\ 
\hline
    Det.       & 34BJ         &      VMC       &    LRDMC\\
\hline
$5s5p$   &  $2s2p$   &  -19.9159(5)   & -19.9422(2)\\
$5s5p$   &  $3s3p$   &  -19.9204(3)   & -19.9433(2)\\
$5s5p$   &  $4s4p$   &  -19.9200(3)   & -19.9423(2)\\
$4s4p$   &  $3s3p$   &  -19.9185(3)   & -19.9418(2)\\
$6s6p$   &  $3s3p$   &  -19.9205(3)   & -19.9430(2)\\
$5s3p2d$ &  $2s2p$   &  -19.9208(3)   & -19.9430(2)\\
$5s5p2d$ &  $3s3p$   &  -19.9213(3)   & -19.9430(2)\\
\hline\hline
\end{tabular}
\end{center}
\end{table}

\subsubsection{\ce{C2}  convergence in the Jastrow basis set}
\begin{table}
\begin{center}
\caption{\small{JHF and JAGPn$=8$ total energies (in 
Hartree) for the \ce{C2} pseudo-molecule. 
Energies were computed at the experimental bond length using
three different basis set for the three-four-body Jastrow 
term.}}
\label{c2_total_energies}
\begin{tabular}{|lccc|} 
\hline\hline 
          &   & VMC & LRDMC \\
\hline
JHF  &(34BJ $2s2p$) & -11.0239(3) & -11.0539(3)\\
JHF  &(34BJ $3s2p$) & -11.0245(2)& -11.0543(2) \\
JHF  &(34BJ $3s3p$) & -11.0250(2)& -11.0550(3) \\
\hline
JAGP$n=8$ &(34BJ $2s2p$)& -11.0535(1) & -11.0748(2)\\
JAGP$n=8$ &(34BJ $3s2p$) & -11.0536(2)&-11.0745(1) \\
JAGP$n=8$ &(34BJ $3s3p$) & -11.0544(1)&-11.0745(2)\\
\hline\hline
\end{tabular}
\end{center}
\end{table}

We further checked the effects of a larger three and four-body Jastrow 
factor (34BJ) in the case of \ce{C2}. We performed simulations
at the experimental bond length with the JHF wave function
and the JAGP with $n=8$  molecular orbitals (whose results 
agrees with the JAGPn$^*$ as shown in Appendix~\ref{c2molapp}) 
with the $2s2p$ 
Jastrow used for all the other cases and for two larger
basis sets (namely $3s2p$, and $3s3p$). Results are shown in 
Table~\ref{c2_total_energies}.  
As expected, effects on the molecular total energy 
of a larger three- and four-body Jastrow
are negligible in the LRDMC at least for the RVB wave 
function. 
For all the other cases, the largest Jastrow basis provides 
an energy gain of at most 1 mH  with respect to the smaller
basis.


\begin{thebibliography}{99}
\bibitem{c2pseudo} We computed the binding energy of the all electron and 'pseudo'- 
\ce{C2}
for a bond length of $2.35$ a.u. and found a LRMDC value of $6.34(2)$ eV and 
$6.298(8)$ eV respectively. 
The discrepancy between the two findings is $1.5(8)$ mH  and is within our accuracy limit.
\bibitem{be2limit} 
Such poor description of the binding energy at experimental bond length 
occurs because the JAGP for the molecule 
is not as accurate  as  the atomic JAGP  (99.2\%  and 99.7\%\cite{casula1} 
of the correlation energy at the DMC level, 
respectively).
\end{thebibliography}
\end{document}